\documentclass[11pt]{article}

\usepackage{amsfonts,amsmath,amssymb,graphicx,bm}

\newcommand{\mycaption}[1]{\caption{\small #1}}

\newcommand{\pr}{\mathbf{P}}
\newcommand{\ex}{\mathbf{E}}

\newcommand{\notthis}[1]{}

\newcommand{\inv}{^{-1}}

\newcommand{\avg}[1]{\overline{#1}}

\begin{document}

\title{\bf Fixed income portfolio optimisation: Interest rates, credit, and the efficient frontier
}

\author{Richard J. Martin\footnote{Department of Mathematics, Imperial College London, Exhibition Road, London SW7 2AZ, UK. Email: {\tt richard.martin1@imperial.ac.uk}}
}
\maketitle

\begin{abstract}
Fixed income has received far less attention than equity portfolio optimisation since Markowitz' original work of 1952, partly as a result of the need to model rates and credit risk. We argue that the shape of the efficient frontier is mainly controlled by linear constraints, with the standard deviation relatively unimportant, and propose a two-factor model for its time evolution.
\end{abstract}

Despite the passage of some 65 years since the emergence of Markowitz's portfolio theory \cite{Markowitz52}, almost all of the discussion on portfolio optimisation has been confined to equity portfolios.
Fixed income is more difficult because there are two main drivers---rates and credit---and unlike equities both have term structures associated with them and are driven by a concept of yield or spread, which influence the construction of expected return.
Credit has the additional feature that the upside is typically far smaller than the downside and so we cannot construct risk on the premise of Normal distributions.
Another differentiator is the use of credit ratings, which in turn are connected to regulatory capital: this too has no counterpart in equities.
Our interest lies both in top-down fixed income portfolio selection from a set of broad asset classes, which we refer to as \emph{sleeves} (for example, US BB~rated corporates), and also bottom-up security selection. Here we mainly consider the first of these problems, but what we say applies equally to both.

The main thrust of recent literature \cite{Korn06,Puhle07,Caldeira12} has been to use yield curve models as a way of understanding risk and return (rather than trying to deal with bond prices directly). We agree with this. (In a recent paper \cite{Deguest18} the authors do not use yield curve models, as ``this reduces misspecification risk''[p.8], but that is a specious argument, as price volatility determines yield volatility and vice versa.)
Remarkably, though, two fundamental issues are entirely absent except in \cite{Deguest18}: duration constraints, and credit risks. We will consider these in depth, but also have another concern: we do not agree that mean-variance is the best framework for understanding fixed income. This is not just because stdev\footnote{Standard deviation.} does not capture tail risks---if this were the objection we would simply recommend CVaR\footnote{Conditional Value at Risk, also known as Expected Shortfall. See for example \cite{Uryasev01} in its application to equity portfolio optimisation.} as a risk measure---but because fixed income portfolios are subject to many different sources of risk, and it is likely to be more effective to explicitly constrain all of these rather than aggregate them into a single quantity.
Put differently, simply knowing the stdev of an instrument or sleeve, and its `beta' to an index, is much less helpful than understanding how it behaves in a variety of interest rate and credit scenarios, and optimisation must address this directly.

Indeed, our central thesis is that fixed income portfolio optimisation is mainly about linear programming (LP), and the construction of risk measures through the maximum of losses in different prescribed rates and credit scenarios, allowing moderate and extreme losses to be constrained.
Notice that we only need \emph{define} these scenarios---it is not necessary to estimate their \emph{probabilities}---whereas to calculate the stdev, CVaR or whatever, their probabilities will need to be known.
Another benefit is simplicity as we do not need to estimate or deal with the covariance matrix of the assets: we simply need the expected value or loss of each asset in each scenario.
Further, for a long-only fixed income portfolio, we scarcely need to constrain stdev, because if the interest rate and credit duration are tightly constrained, the stdev will necessarily be low.

Another matter of interest is the shape of the efficient frontier and how it has evolved over time. 
Many discussions about fixed income asset allocation go no further than drawing a chart of  return vs risk, marking the various asset classes upon it; this is unhelpful to the portfolio manager who must have no more than (say) 15\% in emerging markets, 10\% in European periphery, an average credit quality no worse than BB+, a maximum duration of 6, etc. Such constraints reduce the set of permissible portfolios and hence the maximum return for a given level of risk: therefore they alter the efficient frontier just as much as do market conditions. For a given set of sleeves and constraints, we use a simple new model to directly parametrise the efficient frontier, and show that two easily-interpretable factors capture almost all its variation.

The paucity of research of fixed income portfolio optimisation requires us to spend some time laying the foundations, which we do in the next sections. After that we introduce our model for the efficient frontier and go on to present `live' results.


\section{Interest rate risk and return}
\label{sec:ir}

Tehe natural framework for expressing views on duration and curve risks is that of Heath--Jarrow--Morton \cite{Heath92}, which expresses the motion of the discount-factor curve $B_t(T)$ through the instantaneous forward rate $f_t(T)=-(\partial/\partial T) B_t(T)$, and in particular through its volatility $\sigma$. We refer the reader to \cite[\S17.7]{HullEd3} for a full discussion of what is summarised below.
The quantity $f_t(T)$ evolves as
\[
df_t(T) = \mu(t,T,\Omega_t) \, dt + \sum_i \sigma_i(t,T,\Omega_t) \, dW_{i,t}
\]
in which $W_{i,t}$ are Brownian motions that are not necessarily independent: $\ex[dW_{i,t} \, dW_{j,t}]= \rho_{ij}\, dt$; the symbol $\Omega_t$ denotes all previous history of interest rates that are necessary for determining the forward rate volatility at time $t$. We comment on $\mu$ later. 
\notthis{
The risk-neutral drift of the instantaneous forward rate can be obtained from the forward rate volatility:
\[
\mu(t,T) = \sum_{i,j} \rho_{ij} \sigma_i(t,T,\Omega_t) \int_t^T \sigma_j (t,\tau,\Omega_t) \, d\tau.
\]
} 

In general, the HJM framework does not give rise to Markovian dynamics for the short rate, and while this is not essential it is useful for interpretability and analytical tractability in certain problems. 
However, a special case does\footnote{The form can be more general than this but for reasons of space we omit the details.}, which is:
\[
\sigma_i(t,T) = \sigma_i e^{-\kappa_i(T-t)}.
\]
By performing principal component analysis (diagonalising $\bm{\rho}$) we decompose the motion of $f_t(T)$ into a set of probabilistically independent components $\sigma^\sharp_i(T-t) W^\sharp_{i,t}$, in which each $\sigma^\sharp_i$ is a finite sum of exponential functions. This gives the familiar picture of the first, most important, factor giving a roughly parallel shift, the second causing steepening/flattening, and so on.
Specific interest rate scenarios $\omega_i$ can then be obtained by assigning realisations to each $W^\sharp$, so that $\omega_1$ is a shift higher, $\omega_2$ a shift lower, $\omega_3$ a steepening, $\omega_4$ a flattening, and so on. For strongly nonlinear products such as those with embedded options (e.g.\ callable bonds) we will also need different sizes of rate curve move: for example the sensitivity to a 2\% move will not be twice the sensitivity to a 1\% move\footnote{Similarly for portfolios with derivatives. Seasoned campaigners will remember how much Orange County suffered in 1994 as a result of nonlinear products, after the Fed hiked rates 2\% in short order.}.

\begin{figure}
\centering
\begin{tabular}{l}
\scalebox{0.8}{\includegraphics{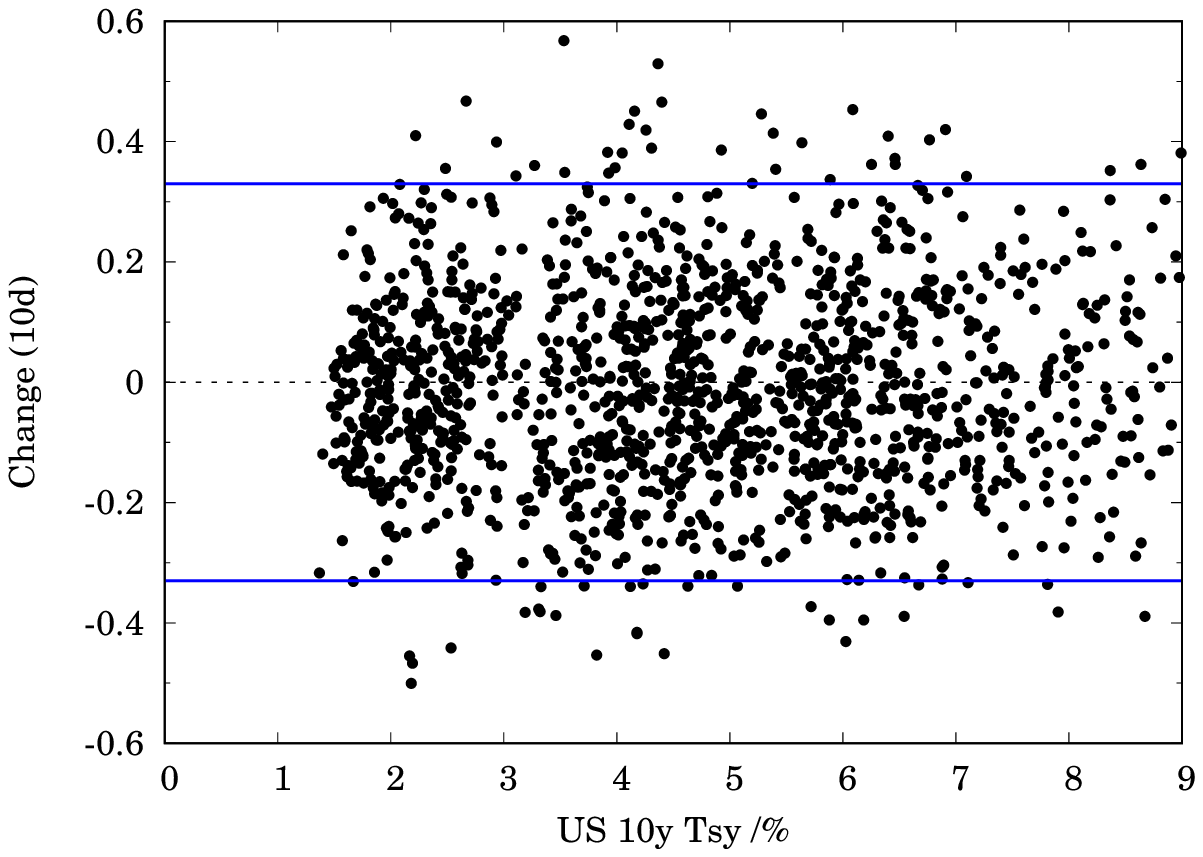}}
\end{tabular}
\mycaption{The magnitude of short-term changes in US 10y Tsy yield has not typically depended on the spot level. Bars are at symmetrical 95\% confidence. Data range 1990--2017. Source: Bloomberg.}
\label{fig:ir}
\end{figure}

The simplest form of this model is when we have only one factor, and for long-only portfolios this will do most of what is needed. By solving for the forward rate and noting that $f_t(t)=r_t$, and then differentiating, the short rate dynamics emerge as 
\begin{equation}
dr_t = \big(\theta(t) - \kappa r_t \big) \, dt + \sigma_r \, dW_t,
\label{eq:hw1}
\end{equation}
which we recognise as the Hull--White model \cite{Hull90}; the function $\theta$ is obtainable from today's term structure. If we make the simplification $\kappa T\ll 1$ then we bump all forward rates by the same amount (parallel bump). 
Apart from analytical tractability there are two advantages to Hull--White. The first is that having $\sigma_r$ is constant, rather than depending on $r$ through a function of the form $\sigma \sqrt{r}$ or $\sigma r$, as seen in the Cox--Ingersoll--Ross and Black--Derman--Toy models respectively, causes no difficulty when rates go negative. The second is that it aligns fairly well with historical experience, as can be seen from the following observation: in the last thirty years the (absolute) volatility of bond yields has not depended strongly on the level of rates.
Figure~\ref{fig:ir} shows this for the typical ten-day move in the 10y Treasury yield. Further, Table~\ref{tab:ir} shows that a Gaussian assumption with a constant yield volatility $\sigma_y=0.90\%$ is remarkably good, though there is the usual caveat that it should not be relied upon for estimating very high levels of confidence, particularly over short time scales.

Some simplifications are possible. 
Rather than using the sensitivity to the instantaneous forward curve (or the zero curve: a parallel shift in one is the same as a parallel shift in the other), simply bump all the bond yields and use the yield-price relationship, which for a bullet bond of maturity $T$ and coupon $c$ paid $m$ times a year is, as a fraction of par,
\begin{equation}
P(T,c;y) = (1+y/m)^{-mT} + c \, \frac{1-(1+y/m)^{-mT}}{y}.
\label{eq:bpx}
\end{equation}
This can be made even simpler if the yield bump is small, by using the duration and convexity. 
While bumping the yield is not the same as bumping the forward rate (because the yield is not a linear function of it), it has the advantage that it is not necessary to build the instantaneous forward curve or the zero curve. This treatment is very common in practice.

For the purposes of this discussion we are most interested in the first interest rate factor and are concerned with the effect of a moderate to large increase in rates, on a 1y horizon. If we regard this as commensurate with a move of a little over two stdevs, and the annual volatility of the USD curve is $\sim0.9$\%, then this is a 2\% rate increase.

The HJM framework establishes the forward rate drift $\mu$ that makes the expected discounted value of a zero-coupon bond a martingale under the risk-neutral measure $\pr_0$. Under a subjective measure $\pr_1$ the drift can be different and can be modelled by regressing the Brownian motions $W_{i,t}$ on an assumed set of technical and econometric factors. Common ideas include: short-term momentum, long-term mean reversion, and aversion in risky asset classes (flight to quality). More simply a common assumption, which we use in the simulation later, is carry-and-rolldown, which is that $f_t(T)$ stays fixed\footnote{Another way of expressing this it to make $f_t(T)$ a $\pr_1$-martingale and ignore any convexity that the traded instruments may have.}: this makes higher-maturity bonds attractive in a steep yield curve environment.

\begin{table}
\centering
\begin{tabular}{lrrrrr}
\hline\\[-12pt]
& $\avg{|\Delta x|}$ & $\avg{(\Delta x)^2}$ &  $\avg{(\Delta x)^4}$ & 95\% & 99\% \\ \hline
Actual & 0.137 & 0.0303 & 0.00360 & $[-0.322,+0.347]$ & $[-0.442,+0.465]$ \\  
$\sigma_y=0.9\%$ & 0.141 & 0.0345 & 0.00289 & $[-0.345,+0.345]$ & $[-0.453,+0.453]$ \\
\hline
\end{tabular}

\mycaption{Statistics of 10-day changes ($\Delta x$) in the  10y Treasury yield, cf.\ Normal distribution.
$\avg{\cdots}$ denotes (time) average.
}
\label{tab:ir}
\end{table}

\section{Credit risk and return}
\label{sec:cr}

\subsection{Risk}

Credit risk depends on the fortunes of all the issuers involved in the portfolio, which is potentially a very high-dimensional problem. A full treatment is not possible in a paper of this size, but we can make some salient observations, and demarcate some problems that are easier than others:
\begin{itemize}
\item[(A)]
A portfolio that is long-only and with no significant issuer concentrations;
\item[(B)]
A market-neutral portfolio;
\item[(C)]
A derivatives trading book with possibly large issuer concentrations and long and short positions across the curves of some or all of the issuers.
\end{itemize}
It is obvious that (A) is the easiest to deal with, and the only one that can be tackled reasonably well using `credit duration'.
To this end we should bear in mind that credit spreads typically move proportionally to the spread level, suggesting \begin{equation}
ds_t =  (\ldots) \, dt + \hat{\sigma} s_t \, dZ_t
\label{eq:cr}
\end{equation}
with $Z_t$ a process of unit quadratic variation ($\ex[dZ_t^2]=dt$). In the simplest setup, $Z_t$ is a Brownian motion and we have a lognormal model.

\begin{table}
\centering
\begin{tabular}{lrrrrr}
\hline\\[-12pt]
& $\avg{|\Delta x/x|}$ & $\avg{(\Delta x/x)^2}$ &  $\avg{(\Delta x/x)^4}$ & 95\% & 99\% \\
Actual & 0.0667 & 0.00891 & 0.000784 & $[-0.142,+0.231]$ & $[-0.237,+0.352]$ \\  
$\hat{\sigma}=40\%$ & 0.0626 & 0.00620 & 0.000120 & $[-0.142,+0.166]$ & $[-0.183,+0.223]$ \\
\hline
\end{tabular}
\mycaption{Statistics of 10-day relative changes ($\Delta x/x$) in the  CDX.IG 5y spread, cf.\ lognormal distribution. $\avg{\cdots}$ denotes (time) average.
}
\label{tab:cr}
\end{table}

\begin{figure}
\centering
\begin{tabular}{l}
\scalebox{0.8}{\includegraphics{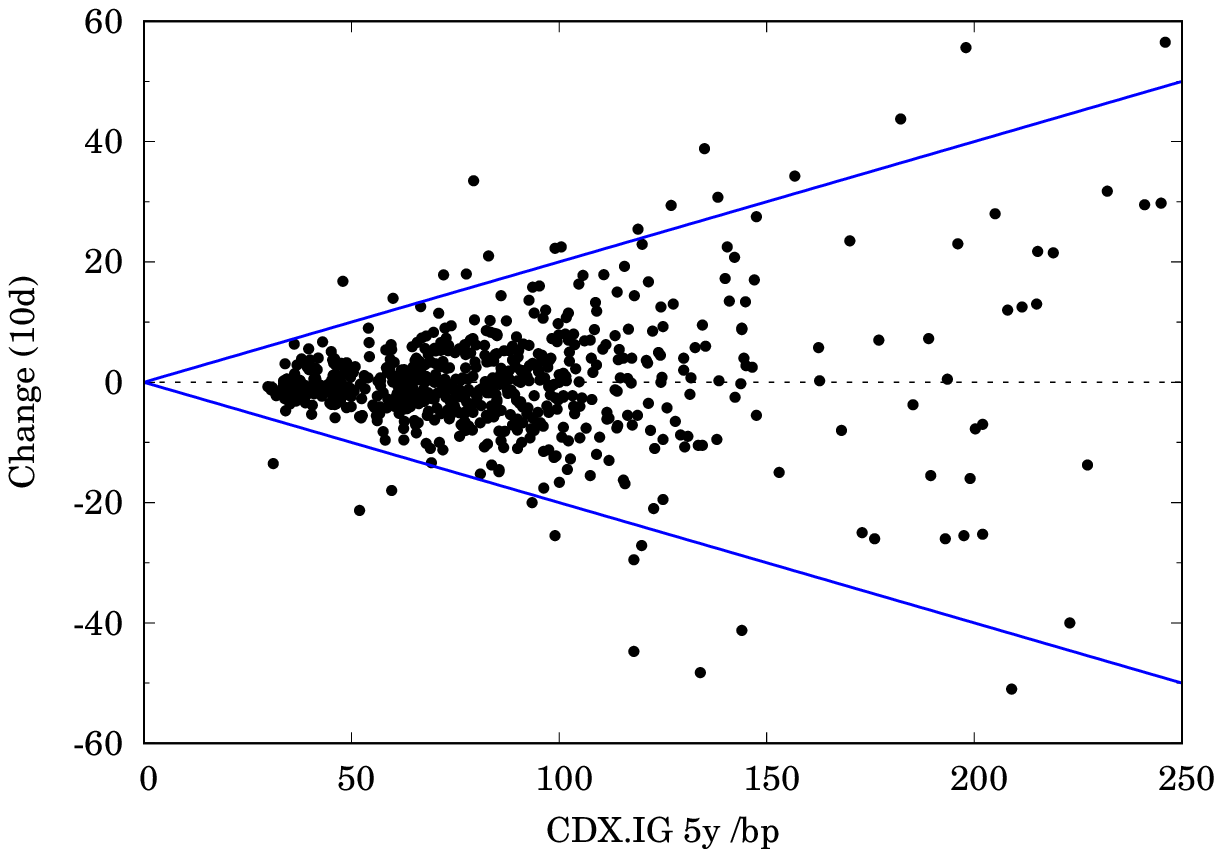}}
\end{tabular}
\mycaption{Moves in credit spread (CDX.IG 5y shown here) are typically proportional to the spot level. Bars are at symmetrical 95\% confidence. Data range 2004--2017. Source: Bloomberg.}
\label{fig:cr}
\end{figure}

For indices, or more generally well-diversified credit pools, a lognormal assumption captures the majority of market moves but does not capture large systemic shocks and is therefore incorrect at high levels of confidence: see Table~\ref{tab:cr} and Figure~\ref{fig:cr} which use $\hat{\sigma}=40\%$. Note that credit index options are typically priced with essentially this model \cite{Martin11c}, and a volatility that is often around 35\%. 
From this comes the well-known idea that we should assess credit spread by bumping spreads proportionally rather than apply a 1bp increase to each credit spread (the so-called CS01), as the higher-yielding ones will typically move more. This gives rise to duration $\times$ spread as a simple measure of risk, and so a large, but not extreme, credit spread increase (by which we understand a little more than 2~stdev), is roughly a doubling of the credit spread: this is a convenient definition which we call CSx2. For extreme moves (credit sress loss, CSL) we propose not relative changes but instead spread \emph{levels} and define the stress loss of an instrument or asset class to be the effect of moving the yield from its current value $y_0$ to a stress level $y_*$. Using (\ref{eq:bpx}) this is
\begin{equation}
\mbox{CSL} = 1 - P(T,y_0;y_*) = (1 - y_0/y_*) \big( 1- (1+y_*/m)^{-mT} \big)
\label{eq:csl}
\end{equation}
and $y_*$ is obtained by adding a specified spread level $z_*$ (Table~\ref{tab:univ}, which is commensurate with the losses sustained in 2008-09) to the riskfree yield of the relevant maturity\footnote{Or LIBOR for floating-rate instruments.}. This kind of constraint operates differently from CSx2 because it is countercyclical, in the sense that as spreads widen, CSx2 rises but CSL \emph{falls}, as losses have already been sustained. Constraining both these measures has the attractive consequence of buying risk when spreads are wide (when CSx2 is the binding constraint), increasing the position as markets rally (as CSx2 falls), but reducing it when they become too tight (as now CSL binds instead).

\begin{figure}
\centering
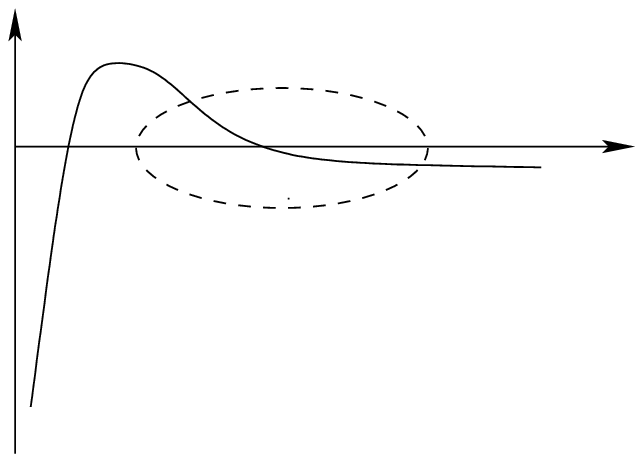
\mycaption{Sketch of PL vs firm value for credit steepener trade discussed in \S\ref{sec:cr}.
A naive argument based on parallel spread curve movements gives information only about the behaviour in the encircled part of the graph.
}
\label{fig:plvsfirm}
\end{figure}

Of the three cases demarcated earlier, (A) is the one that concerns us most here, but it would be remiss not to consider (B) and (C), in which the main difficulty is caused by single names or sectors. 
It is instructive to consider the following problem, set in CDS rather than cash terminology for convenience. Suppose I sell \$50M of 2y protection and buy \$20M of 5y protection on an investment-grade issuer.
As the duration ratio is a little under 5:2, this trade is CS01 positive, which naturally gives the impression of being slightly short risk. The impression is made even stronger if we bump spreads proportionally: as the 5y spread is almost certainly higher than the 2y spread, the trade is predicted to perform well if spreads rise as the 5y spread should move more. But this is only true for \emph{small} moves. If the issuer becomes distressed, the 2y spread will start to rapidly increase so that the curve inverts, and it is obvious that in the event of default there will be a large loss. The relation between the two spreads is complex and a bivariate lognormal distribution does not give the full picture.
The right way to think about this problem is to forget about spreads and yields and return to basics, in the form of the structural (Merton) model, to which an introduction is provided in \cite{Martin07b,Martin10b}. As a function of the firm value, for small moves the trade has a slightly negative delta and positive gamma\footnote{Using option parlance, because we explicitly treat credit as a deep out-of-the money option on the firm.}, whereas for large reductions in firm value the delta becomes positive and the gamma very negative: see Figure~\ref{fig:plvsfirm}.
This motivates the idea that to assess idiosyncratic risk we should reduce the firm value of each issuer by a small amount (say 10\%) and also a much larger amount (say $>50$\%) and constrain the risks involved.
An alternative, which bears strong similarity, is to use rating transition models, which we describe presently, because the credit rating is a proxy for firm value or distance-to-default. As the rating curves for higher-yielding or distressed credits are flat or inverted, the effect shown in Figure~\ref{fig:plvsfirm} is captured.
For `granular' portfolios this may be unnecessary if the issuer concentration is tightly constrained.


\subsection{Expected return}

We cannot simply use credit spread as a proxy for expected return, because it is not guaranteed. 
For a portfolio of individual credits, we in principle require an expected spread change, and default probability, on each name. This is not practicable because each bond has to be updated as the market moves. On the other hand, it is practical to attach a credit rating to each issuer\footnote{Also taking subordination into account where necessary.}, as this only needs to be changed when important news comes out, or there is a change in view on the credit. For these purposes the credit rating does not have to correspond to the public rating: it encompasses our view of the credit quality. With this in mind, there are two key components in the modelling. The first component is a set of credit curves (of spread vs maturity) for each rating, for the sector in question. This can be done using an appropriate numerical fitting procedure such as that discussed in \cite{Martin18c}. The second component is a Markov chain model of transition rates for all ratings from AAA, AA+, \ldots, down to default, see e.g.~\cite{Jarrow97,SandP19}.

If the bond is of maturity $T$ and we are looking at the ER at time horizon $t<T$ then we reprice the cashflows of a $T-t$ maturity bond in each rating state $j$ (say) and weight the results by the transition probability from the current rating $i$ to $j$.
Symbolically if $p_{i\to j}(t)$ denotes the transition probability over time $t$ and $P_j(T,c)$ denotes the price of a bond of rating $j$, maturity $T$ and coupon $c$, then the expected value of the bond at time $t$ is
\[
\sum_{j=1}^D p_{i\to j}(t) P_j(T-t,c)
\]
($D$ denotes the state of default). The total return is obtained by discounting this back to today, adding in the coupon accrued over time $[0,t]$, and subtracting the current price $P_0$. By judicious algebra\footnote{Essentially, shuffling terms and noting that $\sum_{j=1}^D p_{i\to j}(t)=1$ for any $i,t$.} this gives
\begin{equation}
\mathrm{TR} = \left. \begin{array}{l}
\;\; ct + B_0(t) P_i(T-t,c) - P_i(T,c) \\
\mbox{} +  B_0(t) \sum_{j=1}^D p_{i\to j}(t) \big(P_j(T-t,c) - P_i(T-t,c) \big) \\
\mbox{} + P_i(T,c) - P_0 
\end{array} 
\right\},
\label{eq:crer}
\end{equation}
understood as follows.
The first row of terms is the carry and rolldown of a bond rated $i$, assuming the rating and associated spread curve remain fixed. The second adjusts for rating migration and default, and is negative. The third is the cheapness of the bond relative to the curve for rating $i$, so that it is assumed that convergence will occur towards that rating curve over time $t$, as the bond need not currently price in accordance with its rating.
In the interests of simplicity, some approximations have been made: the accrued $ct$  will only be received if the issuer does not default in the period $[0,t]$, so the first term needs to be slightly reduced\footnote{If the hazard rate is $\lambda$ then it should be $(1-e^{-\lambda t})\lambda\inv c < ct$.}; the credit spread of the bond may not converge to its assumed rating curve by time $t$; we may have the view that the curves themselves will move, independently of any rating transition by the credit(s). These issues can be corrected, at the expense of a little extra complexity.

If we ignore the third term then (\ref{eq:crer}) reduces to carry and rolldown minus a hurdle representing the expected loss from rating downgrade and default (and also upgrade but this is not enough to counterbalance the other two).
For high yield credits this hurdle is significant. As an example, assume that a B~rated credit has a 1y default probability of ca.~4--5\%. If the bond is trading at par and recovery is assumed to be 30\%, then a spread of at least 300bp will be needed to justify buying the bond. If the riskfree rate is 2\% then this translates to a yield of around 5\%. Taking into account transitions to B$-$/CCC+/CCC, we conclude that unless the bond yield exceeds ca.~5.75\%, it generates no ER.
Not taking credit losses into account causes the portfolio to become barbelled, as the ER of high-yield assets is greatly overestimated: the optimiser buys these in some proportion and allocates the remainder to cash so as to satisfy whatever average rating constraint is given.

\subsection{Structured credit}

A full discussion of structured credit modelling is not possible but one issue is paramount.
A CDO/CLO tranche rated AA pays a much higher spread than a AA~rated corporate (the latter trade flat to LIBOR).
Anecdotally, the investment world still has trouble explaining this phenomenon, and a commonly-proffered explanations are illiquidity and opacity. While there is some truth in these, they do not address a basic fact: at the senior end of the capital structure, CLOs concentrate market risk, and according to the Capital Asset Pricing Model\footnote{The CAPM in its usual form is not ideally suited to credit, but it is possible to reformulate it using CVaR rather than stdev as a risk measure, and on so doing the conclusion is essentially the same. Except in distressed debt it is hard to find truly idiosyncratic risks in the credit market.} which is the risk for which an investor is compensated.
A portfolio of AA~rated companies is likely to be well diversified, whereas a portfolio of AA CLO tranches is just a systematic bet---all the more so in Europe where different European CLOs are written on substantially the same pools.
The former portfolio is less risky than the latter, so the market cannot price them equally. Such tranches also have a nonlinear response to index widening and so their stress losses (as eq.~(\ref{eq:csl})) are higher than AA corporates.
Failure to incorporate this causes gross overallocation to structured credit.

\section{Optimisation}
\label{sec:opt}

\subsection{General principles}

We define the excess expected return (ER) as the expected return less the riskfree rate, and risk as the value of a risk function which will be defined presently. 
Most problems reduce to:
\begin{equation}
\mbox{Maximise ER s.t.} \left\{
\begin{array}{l}
\mbox{Risk $\le$ limit} \\
\mbox{Other constraints.}
\end{array}\right.
\label{eq:opt0}
\end{equation}
A misconception is that we should attempt to maximise the Sharpe ratio. The table below, with three hypothetical portfolios, shows why this is wrong (ER and risk are annualised, and in context risk means stdev):
\begin{center}
\begin{tabular}{rrrr}
\hline
 & ER & Risk & SR  \\
\hline
1 & 1\% & 0.5\% & 2 \\
2 & 5\% & 4\% & 1.25 \\
3 & 20\% & 25\% & 0.8 \\
\hline
\end{tabular}
\end{center}
In context, Portfolio~1 offers the highest SR but too little ER to be attractive; to make it more so it would need to be levered, which may be expensive or even impossible.
Portfolio~3 offers the highest ER, exhibiting the sort of risk that might reasonably come from a distressed debt portfolio---but in that case one want a much higher ER.
Portfolio~2 offers similar characteristics to a credit portfolio of borderline IG/HY credit quality in a steep yield curve environment, where rolldown and a benign view of credit performance might offer such characteristics.
This sort of risk-return characteristic might be observed \emph{ex~post}, in a year in which markets `went up and up', but as an \emph{ex~ante} statement about expected return and volatility it looks too optimistic, especially in the current market environment. 
It may be fairly said that Portfolio~2 is the most attractive (partly because it is virtually impossible to obtain), despite having neither the highest ER nor the highest SR.

In the context of interest rate risk alone, the following are standard problems which have been considered as mean-variance optimisations (e.g.\ in \cite{Puhle07,Caldeira12}) but are easier to formulate using LP. First, an outright optimisation:
\[
\mbox{ Maximise } {\textstyle \sum_{j=1}^n c_j u_j} \mbox{ s.t. } \left\{ \begin{array}{l}
\sum_{j=1}^n u_j \Delta X_j(\omega_i) \ge -\varepsilon_i \\
0 \le u_j \le \overline{u}_j
\end{array}\right.
\]
where $c_j$ is the ER of the $j$th instrument (we have to take a view on rates, because under $\pr_0$ the ER is always zero), $\Delta X_j(\omega_i)$ is the return of the $j$th bond in the $i$th scenario, $-\varepsilon_i$ is the worst acceptable loss in the $i$th scenario, $u_j$ are the weights to be determined, and $\overline{u}_j$ is the $j$th allocation limit. 
(Transaction costs can easily be incorporated.)
Secondly, index tracking, where risk pertains to the difference between the index and the tracker. For this we can decide whether or not to take a view on rates. If we do then the problem is a variant of the first:
\[
\mbox{ Maximise } {\textstyle \sum_{j=1}^n c_j u_j} \mbox{ s.t. } \left\{ \begin{array}{l}
\sum_{j=1}^n u_j \Delta X_j(\omega_i) - \Delta Y (\omega_i) \ge -\varepsilon_i \\
0 \le u_j \le \overline{u}_j
\end{array}\right.
\]
where $\Delta Y$ is the return of the index in the $i$th scenario. If we do not then we no longer need to know about ER and can minimise the tracking error in the worst scenario:
\[
\mbox{ Maximise } \min_i \left\{ \sum_{j=1}^n u_j \Delta X_j(\omega_i)  - \Delta Y(\omega_i) + \varepsilon_i \right\} \mbox { s.t. } 0 \le u_j \le \overline{u}_j
\]
This minimax problem can be rewritten by introducing a `slack variable' $u_0$ to obtain:
\[
\mbox{ Minimise } u_0 \mbox{ w.r.t.} (u_j)_{j=1}^n \mbox{ s.t. }
\left\{
\begin{array}{l}
u_0 + \sum_{j=1}^n u_j \Delta X_j(\omega_i)  - \Delta Y(\omega_i) + \varepsilon_i  \ge 0 \; \forall i \\
0 \le u_j \le \overline{u}_j
\end{array}
\right.
\]
which is again a standard LP problem. 
If, on solution, $u_0>0$ then it means that it is impossible to make the tracking error as small as is desired.
This same method has successfully been applied to static hedging problems in credit \cite{Martin08a}. Notice that throughout we have constrained the weights $(u_j)$ to be nonnegative, as in each case we are trying to replicate a long-only index. 
Puhle \cite{Puhle07} failed to do this, on the grounds that in the context of a mean-variance problem it leads to solutions that cannot be written in closed form---an entirely spurious objection, as LP problems are so quick to solve---and then ended up with impractical portfolios consisting of delicately balancing long and short positions, a consequence that should have been obvious at the outset\footnote{Aside from this is his disproportionate emphasis on (hypothetical) portfolios of zero-coupon bonds, which represent a tiny proportion of the bond universe---in the context of corporate and sovereign bonds, we estimate $<0.1\%$---which is another reason why his conclusions are of little practical value.}.


Typical constraints encountered in fixed income can be grouped into these categories:

\begin{itemize}

\item
Allocation limits, which for bottom-up models will be primarily sector, country and issuer limits, and for top-down models, sleeve limits;
\item
Rating constraints e.g.\ simply the proportion of sub-IG assets, but also an average rating on a linear scale\footnote{AAA=1, AA+=2, AA=3, AA$-$=4, \ldots}, or Moody's weighted average rating factor (WARF) which is nonlinear and penalises high-yield credits much more\footnote{AA=20, A=120, BBB=360, BB=1350, B=2720, CCC=6500.};
\item
Regulatory capital, e.g.\ NAIC in the US and Solvency Capital Ratio (SCR), which takes duration into account, in the EU; 
\item
Scenario-based risk measures such as IR01, CS01, CSx2, CSL;
\item
Nonlinear risk measures such as stdev, CVaR.
\end{itemize}
We discuss this in detail next.

\subsection{Allocation limits and regularisation}

The purpose of allocation limits is threefold. First, it may be undesirable for reasons of liquidity to buy too much of a certain asset (e.g.\ more than 5\% of a corporate bond issue).
Secondly, any asset or sleeve may suffer an unexpected idiosyncratic accident, which such limits mitigate.
As an example, suppose we are concerned that Turkish assets might drop 10\% (in price terms), and we do not want to lose more than $0.5\%\times$NAV in this eventuality: we constrain Turkey $\le5\%$. 
Thirdly, there is the difficulty of ill-conditioning. Unconstrained problems have solutions that are very sensitive to model inputs. In the context of LP, the optimal point will jump from one corner solution to another as the expected returns are varied, and mean-variance problems suffer from similar problems associated with ill-conditioning of the covariance matrix\footnote{Which are traditionally obviated by Tykhonov regularisation, i.e.\ imposing a small penalty $\propto \|u\|^2$, which is like adding an idiosyncratic risk to each asset.}. 

\notthis{
Constraints must be convex, so that if two portfolios $X_1,X_2$ are feasible\footnote{A portfolio is said to be feasible if it satisfies all the constraints.} then so is any admixture of them. The vast majority of constraints are linear, and indeed all of the ones above are, except stdev and CVaR.
} 

\subsection{Regulatory capital}

In past years the matter of RAROC (ER $\div$ regulatory capital) has been the subject of more attention than is necessary. We explain why it is the wrong thing to optimise. 
Obviously when a particular asset is awarded zero risk weight, optimising RAROC generates nonsense, but even if this pitfall is avoided, it also causes allocation to be skewed towards high-yielding assets whose regulatory capital treatment is (too?) lenient, and it also causes overconcentration.
It is also commonly thought that because an asset has a RAROC less than the desired hurdle, it should not be bought. This is also wrong: it is only necessary for the portfolio as a whole to the have high enough RAROC, with the lower-yielding assets providing diversification.
The correct way to understand regulatory capital is simply as an extra constraint. Unfavourable regulatory capital treatment \emph{is} a valid reason \emph{not to} invest in an asset or asset class; favourable regulatory capital treatment is \emph{not} a good reason \emph{to} invest in it.
Ultimately, economic risk is the risk that is being taken, and so it cannot be ignored.

\subsection{Risk measures}


The last twenty years has seen much attention devoted to an axiomatic theory of risk measures, one of the earliest papers being \cite{Artzner99}. However, the volume of research is out of all proportion to its practical significance. It is almost impossible to find a financial disaster attributable to a `poor choice of risk measure'; invariably, the culprit is an incorrect assessment of the probabilities of bad market moves, which is fundamental to any calculation.
The most egregious example in recent times has been the `London Whale' incident, in which, three years after the Global Financial Crisis, Normal distribution assumptions were still being used to assess the VaR of structured credit products\footnote{See \cite[p.286]{Whale13a}: correspondence of P.~Hagan on 07-Feb-12.}. 

In fact, the necessary axioms for portfolio optimisation can be expressed succinctly: (i) positivity; (ii) 1-homogeneity\footnote{If we scale our positions by some factor, the risk is scaled by the same factor.}; (iii) subadditivity, i.e.\ if $X_1,X_2$ are portfolios then for any $\lambda\in[0,1]$,
\[
R\big(\lambda X_1+(1-\lambda)X_2\big) \le \lambda R(X_1)+(1-\lambda)R(X_2).
\]
With these assumed, we can conclude immediately that in the presence of convex constraints, the efficient frontier must be concave\footnote{If two portfolios $X_1,X_2$ lie on the efficient frontier, then any admixture $\lambda X_1+(1-\lambda)X_2$ is feasible (obeys the constraints); its ER is the weighted sum of the ER's of $X_1,X_2$; and its risk is $\le$ the weighted sum of their risks. Therefore the efficient frontier lies to the left of the chord joining $X_1$ and $X_2$.}. The brevity of this discussion might seem disturbing, but it is justified\footnote{For example we shall not be detained by the fact that standard deviation does not satisfy the so-called monotonicity axiom, see e.g.\ \cite{Artzner99}, which is often deemed necessary. If it is, then the logical conclusion is to dump all of Markowitz's portfolio theory.}.

As is well known, stdev and CVaR obey (i)--(iii), but VaR does not obey (iii). The expected loss in a given scenario (e.g.\ CSx2) obeys (ii),(iii). Finally, so too does the maximum of risk measures obeying (ii),(iii), allowing us to combine several different risk measures $R_k$ into a `total' one $\mathcal{R}$ by
\[
\mathcal{R} = \bigvee_k \alpha_k R_k, \qquad \alpha_k>0
\]
where $\bigvee$ denotes `maximum' and $(\alpha_k)$ are positive constants. 
As in context at least one of the $R_k$ will be positive, $\mathcal{R}$ also obeys (i). 
Notice that constraining $\mathcal{R}(X)$ to be less than or equal to some specified value $\overline{R}$ is identical to constraining $R_k(X) \le \alpha_k\inv\overline{R}$ for each $k$.

\subsection{Iterated linear programming}

If all the constraints are linear then we have a LP problem, for which standard methods are established and fast \cite{NRC}, even for large-scale problems.
Nonlinear optimisation problems, provided they are convex, can be solved iteratively using \emph{cutting-plane methods} \cite{Bertsekas16}, which owe much to Newton-Raphson.
A convex function $R$ necessarily lies above its tangent at any point, so
\begin{equation}
R(u) \ge R(u_*) + (u-u_*)\cdot\nabla R(u_*)
\end{equation}
where $\nabla R$ is the gradient of $R$, and $u_*$ denotes the allocations at the present stage in the optimisation. So, if the upper limit on $R$ is $\overline{R}$, it is necessary for
\begin{equation}
u \cdot \nabla R(u_*)  \le \overline{R} - R(u_*) + u_*\cdot\nabla R(u_*),
\label{eq:Rlin}
\end{equation}
a linear constraint which is added to the constraint set. The optimisation is rerun, ideally using the current point $u_*$ as its starting-point. Then $u_*$ will move, and we add a new constraint of the form (\ref{eq:Rlin}) and keep rerunning until $R(u_*)$ exceeds $\overline{R}$ by no more than an acceptably small amount\footnote{As it will typically not get below $\overline{R}$ in finitely many iterations. Alternatively, in (\ref{eq:Rlin}) replace $\overline{R}$ by a tighter constraint $\overline{R}-\varepsilon$ and stop when $R(u_*)<\overline{R}$.}.
Note that it is important to add a new constraint each time, rather than moving (\ref{eq:Rlin}) around each time $u_*$ changes. The only requirement is that $R$ be convex and that we can easily evaluate\footnote{In effect the so-called risk contribution, see e.g.~\cite[\S3]{Martin11b}.} $\nabla R$.

\section{Modelling and Results}

\subsection{Efficient frontier model}

The efficient frontier demarcates the maximum level of ER ($r$) for each level of risk ($R$).
It is an upward-sloping, concave function and, assuming that a portfolio with 100\% cash is feasible, it must pass through the origin\footnote{One might enquire why, if all the constraints are linear, the efficient frontier is not simply a straight line. The answer is that at different points on the frontier, different constraints are binding; therefore, the frontier is piecewise linear.}.
Given that risk and return have different units\footnote{If we change the time horizons on which risk or return are calculated, the axes will be stretched accordingly.} the functional form $\psi(r/a,R/b)=0$ is necessary, so it is then a question of picking a sensible function $\psi$.
With this in mind, and considering the required shape, an obvious idea is 
\begin{equation}
r = a_t(1-e^{-R/b_t}), \qquad a_t,b_t>0 ;
\label{eq:efmodel}
\end{equation}
we write $_t$ because over time the parameters are expected to change.
The interpretation of $a_t,b_t$ is that $a_t$ is the ER for a high-risk portfolio, as it is the $R\to\infty$ asymptote, and $a_t/b_t$ is the ER per  unit risk for a low-risk portfolio, as it is the gradient of the efficient frontier at the origin.

In different market scenarios and with different constraints binding, $a_t,b_t$ will behave differently:
\begin{itemize}
\item[(a)]
The riskfree curve steepens, with the IR01 constraint binding. (Note that a parallel shift will have no effect, if a static interest-rate volatility is assumed, because by ER we always mean relative to the front end of the curve.)
Here $a$ fwill increase but not $b$, and the frontier is simply stretched upwards.

\item[(b)]
Credit spreads increase, CS01 constraint binding.
This is the same as (a).
\item[(c)]
Credit spreads increase, with the CSx2 constraint binding.
Here the return and risk increase by the same factor, and the efficient frontier is stretched upwards and to the right so that $a$ and $b$ both increase.
\item[(d)]
Credit spreads increase, CSL binding. Here the ER increases but risk \emph{decreases}, so that $a$ moves up and $b$ down.
\item[(e)]
`Risk off'/`Flight-to-quality'. If yields compress in low-risk portfolios but decompress in high-risk ones, the main effect is that $b$ increases more than $a$, so that $a/b$ (the gradient of the frontier at the origin) decreases.
\end{itemize}
In view of this we can expect that over time $a_t,b_t$ will not be perfectly correlated, so that we will need both to explain the full evolution of the efficient frontier.
Plausibly, they are mean-reverting over a long enough time scale, and so a reasonable econometric model is that $(\ln a_t, \ln b_t)$ follows a bivariate Ornstein--Uhlenbeck process.

\subsection{Simulations}

We construct a risk function defined as
\begin{equation}
\mathcal{R} =  R_1 \vee R_2 \vee R_3 \vee 2R_4
\label{eq:R1234}
\end{equation}
where the risk measures are defined as:
\begin{center}
\begin{tabular}{rl}
$R_1$ & Loss from riskfree rate +2\% (for floating-rate instruments this will be zero) \\
$R_2$ & Loss from credit spreads doubling (CSx2) \\
$R_3$ & Stress losses (CSL) as eq.(\ref{eq:csl}) \\
$R_4$ & Annual stdev \\
\end{tabular}
\end{center}
and the stdev is constructed as follows. Interest and credit risks are assumed uncorrelated, so thst the variances simply add. As discussed earlier, interest rate variance uses a yield volatility of $\sigma_y=0.9\%$ across all maturities, while the relative credit spread volatility is $\hat{\sigma}_c=0.35\%$. The credit risks of different sleeves are correlated with coefficient $\rho=0.8$ throughout.
The portfolio variance $\Sigma^2$ over a time horizon $\Delta t$ (taken as 1y) is therefore
\begin{equation}
\textstyle
\Sigma^2 = \left\{
\big(  \sum_j u_j D^\mathrm{ir}_j \big)^2 \sigma_y^2  + 
\rho \big( \sum_j u_j D^\mathrm{cr}_j s_j \big)^2 \hat{\sigma}_c^2  +
(1-\rho) \sum_j u_j^2 (D^\mathrm{cr}_j)^2 s_j^2 \hat{\sigma}_c^2   
\right\} \Delta t
\end{equation}
where $D^\mathrm{ir}_j$ and $D^\mathrm{cr}_j$ are the interest rate and credit duration of the $j$th sleeve.

We then vary the total risk limit $\overline{R}$ and plot the ER that the optimisation finds, and also fitting (\ref{eq:efmodel}). This is done on different dates going back to 2000 in 3m steps.


The asset classes are described in Table~\ref{tab:univ}, which also shows the Bloomberg tickers for the yield\footnote{Historical spread and total return information is available for these sleeves, but for reasons of space is not shown.} and the allocation limits imposed on each sleeve.
They are: US~Treasuries of various maturities, US investment grade and high yield corporate bonds ranging from AA to CCC; US and European subordinated financials (the latter being swapped into USD); emerging market corporate/sovereign IG/HY; residential mortgage-backed securities; CLOs from AAA to BBB. 
The allocation constraints on the various sectors are: high-yield (of all types) $\le 60\%$; emerging markets $\le 15\%$; structured credit $\le 10\%$; financials $\le 20\%$. The only other constraint is that the average credit rating be no worse than BB.

\begin{figure}[!htbp]
\centering
\begin{tabular}{rl}
(a) &
\scalebox{0.8}{\includegraphics{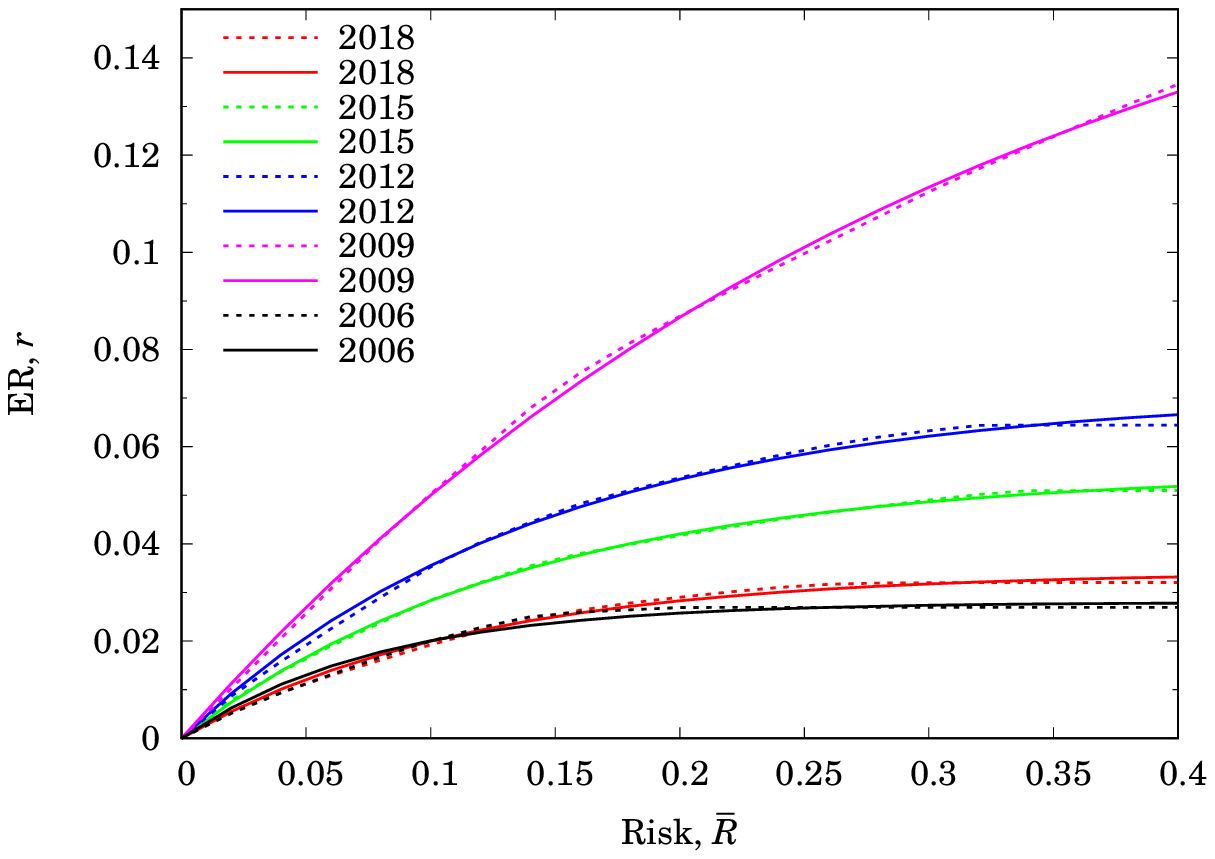}} \\
 &
\scalebox{0.8}{\includegraphics[trim=0 15mm 0 15mm, clip]{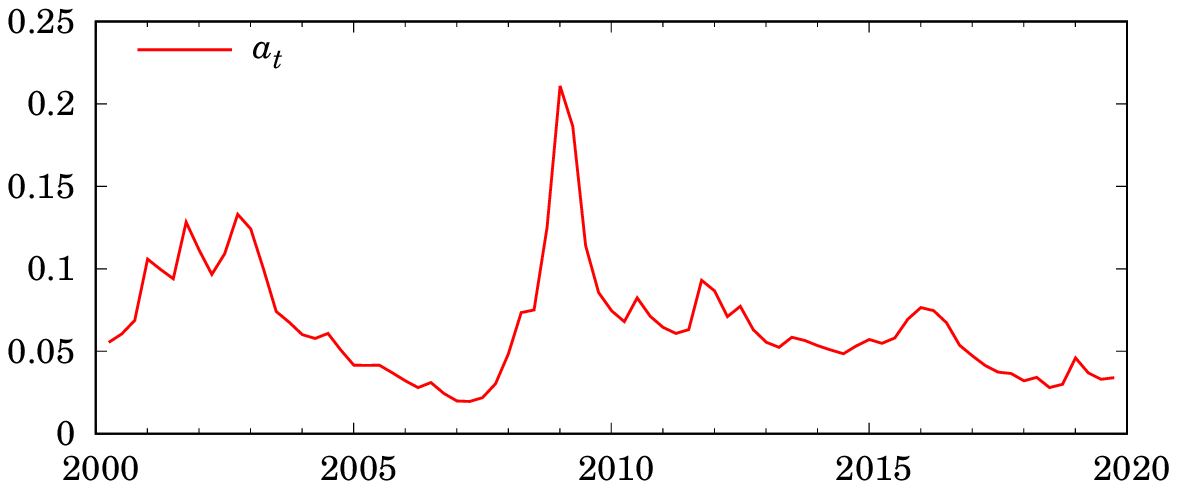}} \\
 &
\scalebox{0.8}{\includegraphics[trim=0 15mm 0 15mm, clip]{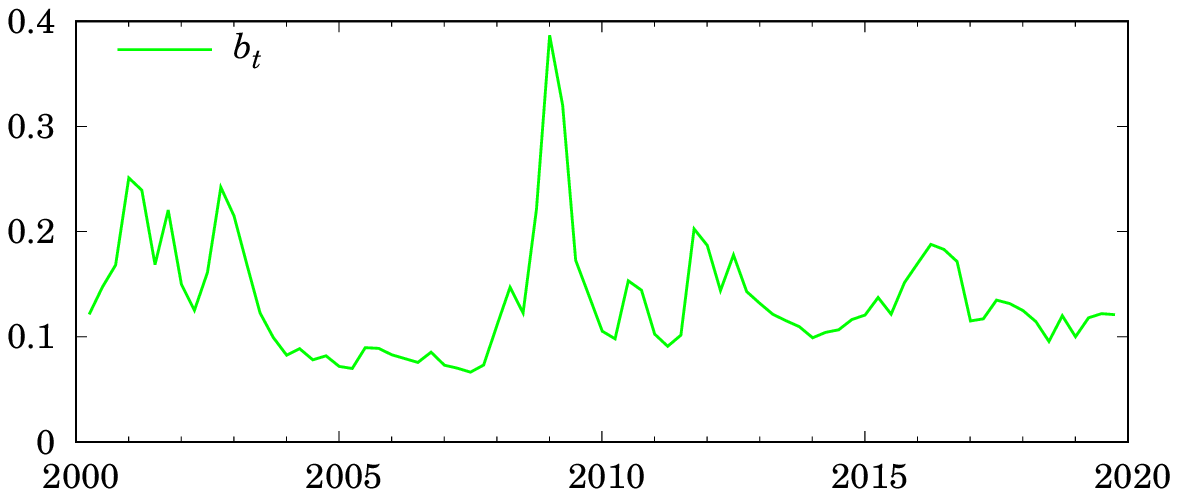}} \\
(b) &
\scalebox{0.8}{\includegraphics[trim=0 15mm 0 15mm, clip]{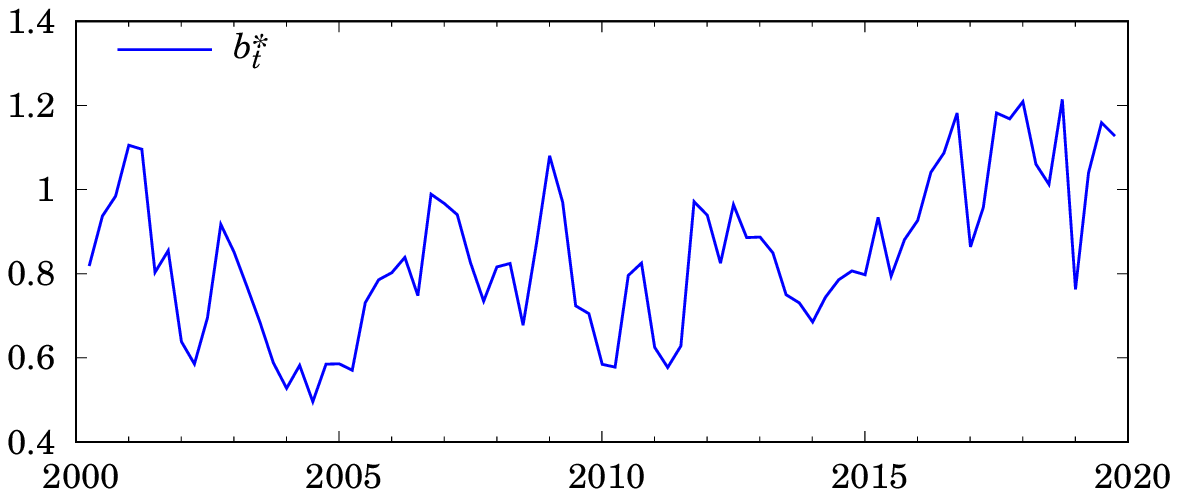}} \\
\end{tabular}
\mycaption{Results for test portfolio. (a) Efficient frontiers (dotted lines) and model fit (solid lines) at beginning of April on dates shown. Units are fractions of portfolio NAV, on both axes.
(b) Time evolution of the factors $a_t$, $b_t$ and second independent factor $b^*_t=b_t/a_t^{0.66}$.
}
\label{fig:2}
\end{figure}

Figure~\ref{fig:2}(a) shows the evolution of the efficient frontier since 2006, and clearly (\ref{eq:efmodel}) is an excellent fit in a variety of different market scenarios. Figure~\ref{fig:2}(b) shows the evolution of $a_t$ and $b_t$, which are seen to be positively correlated.
Independent factors $a^*_t$, $b^*_t$ are obtained by taking $a^*_t=a_t$ as the first and $b^*_t = b_t/a_t^p$, for suitably-chosen $p$, as the second. 
We found $p\approx0.66$ by regression of $\ln b_t$ against $\ln a_t$, and this is shown in Figure~2(c). Their interpretation is that $a^*_t$ represents overall risk appetite, so that as the market goes risk-on, $a^*_t$ declines; whereas $b_t^*$ indicates flight-to-quality having controlled for $a_t$, so that as the market moves into safer portfolios $b^*_t$ increases.
Comparison between 2006 and 2018 is interesting because Figure~\ref{fig:2}(a) shows the efficient frontiers to be of different shape, and indeed the factor $b^*_t$ is higher now than then.

We then run the test without the stdev constraint, so the $2R_4$ term in (\ref{eq:R1234}) is removed.
The results are not identical, but can scarcely be distinguished from Figure~\ref{fig:2} and so we have not plotted them. In view of what we have said before about long-only portfolios, the stdev constraint is largely redundant once the interest rate and credit duration are constrained.


\section{Conclusions}

We have established a framework for doing fixed income portfolio optimisation.
The efficient frontier is mostly a product of linear constraints, and for long-only porfolios the standard deviation constraint is largely unnecessary.
By directly modelling the efficient frontier we can make statements about risk and reward that pertain directly to the portfolio with full regard to the constraints.
The factors $a_t,b^*_t$ almost completely describe how the efficient frontier of a particular portfolio has evolved over time, and give a manager an indication of how risk is being rewarded within the operating constraints.

\notthis{
A possible explanation is that (\ref{eq:efmodel}) is universal in the following sense. The optimisation for the asset allocations $x$ is
\[
\mbox{Maximise } c\cdot x \mbox{ s.t. } \left\{
\begin{array}{l}
Rx \le \overline{R} \\
0 \le x \le \overline{x} \\
Ax \le b
\end{array}\right.
\]
where the lines on the RHS are the constraints: respectively, risk, individual allocations, and `others'.
As we have seen, raising and lowering  $\overline{R}$ causes $c\cdot x$ to trace out the efficient frontier. 
Now suppose that the vector $c$ and matrices $A,R$ are chosen randomly, with the important proviso that their elements be non-negative. Consider an average over all realisations. Does random matrix theory predict (\ref{eq:efmodel})? An excellent review of the background and applications is provided by Bouchaud \& Potters \cite{Bouchaud09}, but the subject is not for the faint-hearted.
} 

\section*{Acknowledgements}

I thank Yang Zhou and Yao Ma for their part in the early development of this work, and Michael Story, Erik Vynckier and John Zito for helpful discussions.

\bibliographystyle{plain}
\bibliography{}

\clearpage

\begin{table}
\centering

\begin{small}
\begin{tabular}[t]{llrrrr}
\hline
Name & Ticker & Max (\%) & Maturity  & $z^*$ (bp) & Float \\ 
\hline
Cash & & 100 & \\
Tsy (5y, 10y, 30y) & CMAT05Y etc. & 100 & 5, 10, 30 & - \\
US Corporate AA &	JULIAAY & 10 & 10 & 100 \\
US Corporate A &	JULIAY	 & 15 &	10 & 200 \\
US Corporate BBB &	JULIBBBY & 20 & 10 & 350 \\
US Corp.~IG 5y &	 IBOXUMAE & 20  & 5 & 250 & $\ast$ \\
CLO AAA &	JCLOAAAY & 5 & 10 & 350 & $\ast$ \\
CLO AA &  JCLOAAYL &	5 & 10 & 550 & $\ast$ \\
CLO A	& JCLOAYLD & 5 & 10 & 900 & $\ast$ \\
CLO BBB &  JCLOBBBY & 5 & 10 & 1700 & $\ast$ \\
US Corporate BB &	JPDKBB	& 5 & 7 & 900 \\
US Corporate B &	JPDKB & 5 & 7 & 1500 \\
US Corporate CCC &	JPDKCCC & 5 & 7 & 3000 \\
US HY 2--4y &	JPDK2/4 & 5 & 3 & 1500 \\
EM Corporate IG &	JBDYIGYW & 10 & 8 & 450 \\
EM Corporate HY &	JBDYHYYW & 5 & 7 & 1100 \\
EM Sovereign IG &	JPBYIGYW & 10 & 12 & 350 \\
EM Sovereign HY & 	JPBYHYYW & 5 & 10 & 1200 \\
Agency real estate &	LUMSYW &	10 & 5 & 1200 \\
US IG Financials &	JULIFINY & 10 & 12 & 500 \\
US HY Financials &	JPDKFINL & 5 & 8 & 1000 \\
EU Insurance (Tier 2)	& JPSUIBEY & 10 & 10 & 700 \\
\hline
\end{tabular}
\end{small}

\mycaption{Universe used in simulation. `Float' denotes LIBOR-based instruments with no interest rate risk. 
Ticker refers to Bloomberg. 
IBOXUMAE is the ticker for the on-the-run CDX.IG~5y CDS index.
}
\label{tab:univ}
\end{table}

\notthis{
\begin{figure}[!htbp]
\centering
\begin{tabular}{rl}
(a) &
\scalebox{0.8}{\includegraphics{effront_fig3a.eps}} \\
 &
\scalebox{0.8}{\includegraphics[trim=0 15mm 0 15mm, clip]{effront_fig3b.eps}} \\
&
\scalebox{0.8}{\includegraphics[trim=0 15mm 0 15mm, clip]{effront_fig3c.eps}} \\
(b) &
\scalebox{0.8}{\includegraphics[trim=0 15mm 0 15mm, clip]{effront_fig3d.eps}} \\
\end{tabular}
\mycaption{As for Figure~\ref{fig:2} but without the stdev constraint. 
}
\label{fig:3}
\end{figure}

\begin{figure}[!htbp]
\centering
\begin{tabular}{rl}
(a) &
\scalebox{0.8}{\includegraphics{effront_fig4a.eps}} \\
 &
\scalebox{0.8}{\includegraphics[trim=0 15mm 0 15mm, clip]{effront_fig4b.eps}} \\
 &
\scalebox{0.8}{\includegraphics[trim=0 15mm 0 15mm, clip]{effront_fig4c.eps}} \\
(b) &
\scalebox{0.8}{\includegraphics[trim=0 15mm 0 15mm, clip]{effront_fig4d.eps}} \\
\end{tabular}
\mycaption{As for Figure~\ref{fig:2} but for Portfolio~B ($b^*_t=b_t/a_t^{0.87}$).
}
\label{fig:4}
\end{figure}

} 

\end{document}